# Sales Prediction Model
# Using Classification Decision Tree Approach
# For Small Medium Enterprise
# Based on Indonesian E-Commerce Data

Raden Johannes, Andry Alamsyah
School of Economic and Business, Telkom University

**Abstract**

The growth of internet users in Indonesia gives an impact on many aspects of daily life, including commerce. Indonesian small-medium enterprises took this advantage of new media to derive their activity by the meaning of online commerce. Until now, there is no known practical implementation of how to predict their sales and revenue using their historical transaction. In this paper, we build a sales prediction model on the Indonesian footwear industry using real-life data crawled on Tokopedia, one of the biggest e-commerce providers in Indonesia. Data mining is a discipline that can be used to gather information by processing the data. By using the method of classification in data mining, this research will describe patterns of the market and predict the potential of the region in the national market commodities. Our approach is based on the classification decision tree. We managed to determine predicted the number of items sold by the viewers, price, and type of shoes.

## 1. Introduction

In the scope of Asia, Indonesia has the 4th rank on internet users. Indonesia has 71.2 million internet users (internet world stats, 2015). This phenomenon proves that internet usage in Indonesia is getting higher and Internet has been an important need. In 2015 e-commerce in the global market is predicted to grow up to 20%: (aprisindo.or.id). The rapid growth of mobile, internet, and population have contributed to e-commerce growth. Tokopedia.com is the largest online marketplace in Indonesia (Lukman, 2014). Indonesian shoe product has reached US $ 3.8 billion and the next year is US $ 4.5 billion sales, that increased by 18.4 %. The trend of keyword *sepatu* on google trend shows the growth of footwear search in Indonesia (Widjanarko, 2015).

Data mining is a discipline that can be used to gather information by processing the data. By using the method of classification in data mining, this research describes patterns of the market and predicts the potential region of the national market commodities. However, research which is a concern on small-medium enterprise with data mining is not widely used in Indonesia. Using a classification decision tree, the seller can boost sales.

## 2. Theoretical Background

A. Social Computing

Social computing is concerned with the study of social behavior, and social context based on computational systems Social computing provides four main facilities to the behavioral modeling (Dubey et al., 2014:1):

1. Model

Building To create & build up models for behavior

**2. Analysis**
Review the creation & already created models with their design work
**3. Pattern mining**
Minimize the patterns through mining
**4. Prediction**
Follow the rules & regulations to control the error in the designing

B. Data Mining

Data mining is the process of discovering insightful, interesting, and novel patterns, as well as descriptive, understandable, and predictive models from large-scale data (Zaki & Meira : 2014). Data mining is widely used in many domains, such as retail, finance, telecommunication, and social media (Zhao: 2012).

C. CRISP-DP

The Cross-Industry Standard Process for Data Mining (CRISP-DM) is a popular methodology for increasing the success of DM projects. This methodology defines a non-rigid sequence of six phases, which allow the building and implementation of a DM model to be used in a real environment, helping to support business decisions (Chapman in Moro et al; 2014).

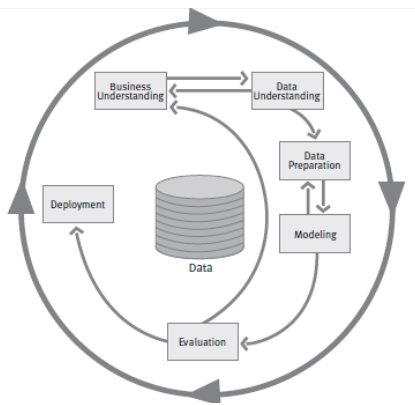

Figure 1. CRISP-DM Process (source: zhao 2012)

D. Classification

Classification models describe data relationships and predict values for future observations (Swamy and Hanumanthappa 2012:1). Classification maps data into predefined groups of classes. It is often referred to as supervised learning because the classes are determined before examining the data (Nancy et al., 2013:583).

E. Decision Tree

The decision tree is a logical model represented as a binary (two-way split) tree that shows how the value of a target variable (output) can be predicted by using the values of a set of predictor variables (input) (Swamy and Hanumanthappa, 2012:1).

F. Web Mining

Web data mining applies data mining technology to the Web and abstracts interesting information from Web text, Web information, Web data, and other Web services. Web data mining has three parts; content mining, user visit mode mining, and structure mining (Lei, 2013:232-233).

G. E-Commerce

E-commerce is a business activity that associate consumer, manufacturing service provider and distributor, using a computer network called the internet (Iyas, 2011:21)

## 3. Methodology

The researcher uses the data mining method to conduct Knowledge Discovery in Database (KDD), KDD is the nontrivial process of identifying valid, novel, potentially useful, and ultimately understandable patterns in data. In data mining, a decision tree is a predictive model which can be used to represent both classifiers and regression models (Milana and Abadyo, 2013).

We investigate data of shoe sales in Indonesia using the web mining method on online marketplace website (tokopedia.com) until March $3^{rd}$, 2015. The attribute of this research are price, type of shoes, insurance, product viewer, city of the seller, rating of speed, accuracy, and service have impacts on product sales. Data used in this research are data of shoe sales in Indonesia which captured using web mining methods on online marketplace websites (tokopedia.com) until March 3rd, 2015. The attribute used in this research are price, type of shoes, insurance, product viewer, city of the seller, rating of speed, service, and accuracy has an impact on product sales.

There are several steps to make a decision tree using the CHAID algorithm which are:

1. Merging
Category merging can be done on an independent variable that has more than two categories that are related.

2. Splitting
In this part independent variable which used as the best split node. Splitting was conducted with a p-value on each independent variable.

3. Stopping
The decision tree should be terminated by the rules. If there is no significant independent variable or if a tree reaches a maximum value limit of the tree defined specifications.

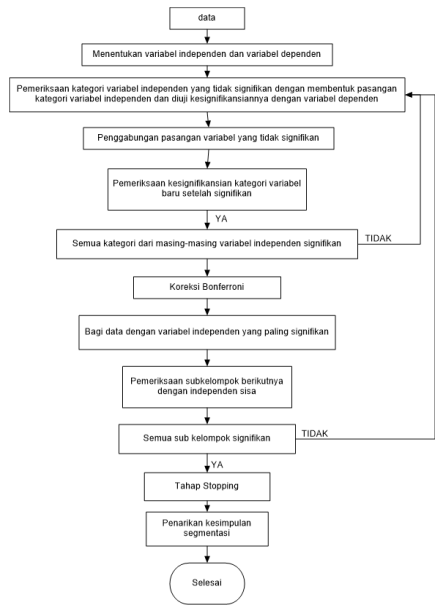

Figure 2. CHAID Algorithm Flowchart

In CHAID analysis, the following are the components of the decision tree:

*Root Node*: *The root node* contains the *dependent variable* or *target* variable. For example, a bank can predict the credit card risk based upon information like age, income, number of credit cards, etc. In this example, credit card risk is the target variable and the remaining factors are the predictor variables.

*Parent's Node*: The algorithm splits the target variable into two or more categories. These categories are called parent nodes or initial nodes. For the bank example, high, medium, and low categories are the *parent's nodes*.

*Child Node*: Independent variable categories which come below the parent's categories in the CHAID analysis tree are called the child node.

*Terminal Node*: The last categories of the CHAID analysis tree are called the *terminal node*. In the CHAID analysis tree, the category that is a major influence on the dependent variable comes first and the less important category comes last. Thus, it is called the terminal node.

*Bonferroni Correction* is a process that occurs when multiple statistical tests for freedom or lack of freedom performed at the same time. *Bonferroni Correction* uses this formula (Permana, 2011):

1. Monotonik Independent Variable

$$M = \binom{c-1}{r-1}$$

2. Independent Variable

$$M = \sum_{i=0}^{r-1} (-1)^i \frac{(r-1)^c}{i!(r-i)!}$$

3. Float Independent Variable

$$M = \binom{c-2}{r-2} + r\binom{c-2}{r-1}$$

Descriptions :
M = *Bonferroni Mining*
c = Total first independent variable
r = Total independent variable after merging

## 4. Data Processing and Modeling

The classification method can be used to predict sales. Classification is one of the most frequently studied problems by data mining and machine learning (ML) researchers. It consists of predicting the value of a (categorical) attribute (the class) based on the values of other attributes (the predicting attributes (Bhardwaj and Pal 2012:1)

The decision tree in figure 3 is the result of tokopedia's data classification. The tree has eleven nodes with seven terminal nodes. The depth of the classification tree is three. Significant variables to form the model are "dilihat", "harga" and "tipe". Each terminal node can be referred to as one of the segments of the "sold" variable. The conclusion of "penjualan" segment can be seen from the percentage interval category of the "sold" variable of each node.

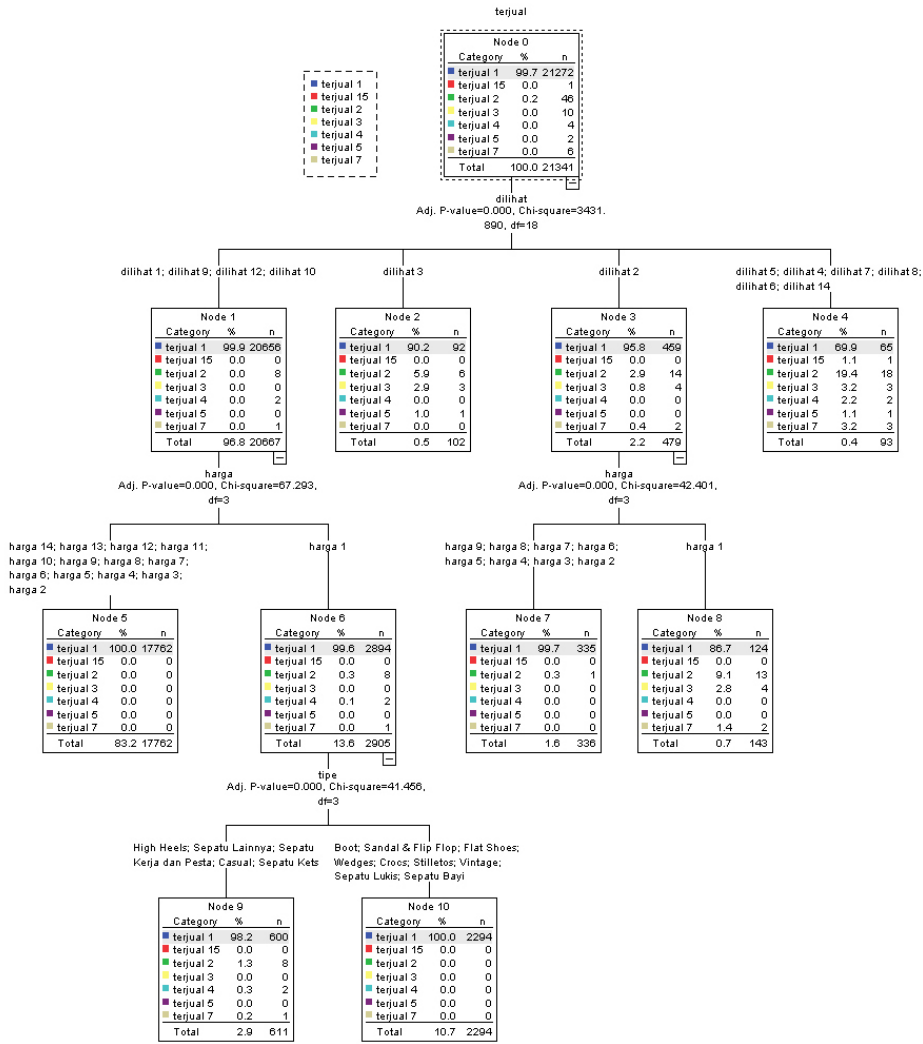

Figure 3. Training Set Decision Tree

TABLE I.  TERMINAL NODES

| Terminal Nodes | Descriptions |
| --- | --- |
| Terminal Nodes 1 | Variable "terjual" that has variable "dilihat" on interval 1, 9, 10, and 12. Then variable "terjual" that has variable "harga" on interval 2, 3, 4, 5, 6, 7, 8, 9, 10, 11, 12, 13, and 14. |
| Terminal Nodes 2 | Variable "terjual" that has variable "dilihat" on interval 1, 9, 10, and 12. Then variable "terjual" has variable "harga" on interval 1. Then variable "terjual" that has variable "type" High Heels, Other, Office Footwear and Sneakers type. |
| Terminal Nodes 3 | Variable "terjual" that has variable "dilihat" on interval 1, 9, and 10. Then Variable "terjual" that has variable "harga" on interval 1. "terjual" that has variable Boot, Sandals & Flip Flop, Flat Shoes, Wedges, Stilettos, Vintage, Painted Shoes and Baby shoe type. |
| Terminal Nodes 4 | Variable "terjual" that has variabel "dilihat" on interval 3 |
| Terminal Nodes 5 | Variable "terjual" that has variable "dilihat" on interval 2. Then variable "terjual" that has variable "harga" on interval 2, 3, 4, 5, 6, 7, 8, and 9. |
| Terminal Nodes 6 | Variable "terjual" that has variable "dilihat" on interval 2, Then variable "terjual" that has variable "harga" on interval 1. |
| Terminal Nodes 7 | Variable "terjual" that has variable "dilihat" on interval 4, 5, 6, 7, and 8. |

With terminal and classification patterns, we can predict sales. For illustration, if we sell shoes at Rp 62.000 with 80 viewers and classified as "Sneakers", there is a 98,2% probability that the shoes will be sold 1-143 unit, 1,3% probability that the shoes will be sold 155-287 unit, 0,3% probability that the shoes will be sold 431-573 unit, and 0,2% probability that the shoes will be sold 860-1002 unit.

## 5. Conclusion

West Java, Jakarta, and East Java are the first three largest locations of footwear sales. When associated with sales count and viewer, we can see that West Java, Jakarta, and East Java are still the first three of these categories. Table 4.2 also shows the number of footwear types based on the city location. The average footwear price on tokopedia.com is Rp. 145.921. The province of Bangka-Belitung Islands has the highest average price, and then South Kalimantan and West Sumatra follow behind. Sales pattern on tokopedia.com shows that the higher the price the higher the sales. The footwear type highest price is the boot, office footwear, and stilettos. And casual footwear has the highest sales count and product views on tokopedia.com. The classification method can be used to predict sales. The decision tree in Figures 4.11 and 4.12 is the result of tokopedia's data classification. The tree has eleven nodes with seven terminal nodes. The depth of the classification tree is three. Significant variables to form the model are "view", "price" and "type". Each terminal node can be referred to as one of the segments of the "sold" variable. The conclusion of the "sold" segment can be seen from the percentage interval category of the "sold" variable of each node.

## References

Bhardwaj, B. K., & Pal, S. (2012). Data Mining: A prediction for performance improvement using classification. arXiv preprint arXiv:1201.3418.
Dubey, A. K., Sisodia, D., Khunteta, D., Saini, A. R., & Chaturvedi, V. (2014). Emerged Computer Interaction With Humanity: Social Computing. International Journal on Computational Sciences & Applications (IJCSA) Vola, (1).
Iyas. (2011). Klasifikasi Data Karyawan Untuk Menentukan Jadwal Kerja Menggunakan Metode Decision Tree.
Laudon, K. C., & Traver, C. G. (2012). E-commerce 2012: Business. Technology, Society.
Lei, Y. (2013). Web Data Mining Technology and Instrument Research. In Proceedings of the 2nd International Conference on Green Communications and Networks 2012 (GCN 2012): Volume 2 (pp. 231-237). Springer Berlin Heidelberg.

Liu, B. (2007). Web data mining: exploring hyperlinks, contents, and usage data. Springer Science & Business Media.

Maimon, O., & Rokach, L. (2008). Data mining with decision trees: theory and applications

MILANA, N. (2013). CHAID UNTUK MENGKLASIFIKASI STATUS MAHASISWA SETELAH LULUS PERKULIAHAN (Studi Kasus Pada Alumnus Prodi Matematika. Jurusan Matematika. FMIPA. Universitas Negeri Malang. Tahun 2007-2012). SKRIPSI Jurusan Matematika-Fakultas MIPA UM.Zhao, Y. (2012). R and data mining: Examples and case studies. Academic Press.

Moro, S., Laureano, R., & Cortez, P. (2011). Using data mining for bank direct marketing: An application of the crisp-dm methodology.

Permana, H. (2011). Klasifikasi dengan Metode CHAID (Chi-Squared Automatic Interaction Detection) dan penerapannya pada Klasifikasi Alumni S1 FMIPA UNY (Doctoral dissertation, UNY).

Swamy, M. N., & Hanumanthappa, M. (2012). Predicting academic success from student enrolment data using decision tree technique. Int. J. Appl. Inf. Syst, 4, 1-6.

Zaki, M. J., & Meira Jr, W. (2014). Data Mining and Analysis: Fundamental Concepts and Algorithms. Cambridge University Press.

Zhao, Y. (2012). R and data mining: examples and case studies. Academic Press.